\documentclass[twocolumn,showpacs,preprintnumbers,amsmath,amssymb,revsymb,pra, superscriptaddress]{revtex4}
\usepackage{graphicx}
\usepackage{mathptmx}
\usepackage{bm}

\newcommand{\D}[2]{\frac{d #1}{d #2}}
\newcommand{\Dp}[2]{\frac{\partial #1}{\partial #2}}

\newcommand{\lbrla}{\left\langle }
\newcommand{\rbrra}{\right\rangle }
\newcommand{\lbrp}{\left| }
\newcommand{\rbrp}{\right| }
\newcommand{\bra}[1]{\lbrla #1 \rbrp}
\newcommand{\ket}[1]{\lbrp #1 \rbrra}
\newcommand{\braket}[2]{\lbrla #1 | #2 \rbrra}

\begin{document}
\title{Time operators in stroboscopic wavepacket basis and the time scales in tunneling}
\author{P. Bokes } \email{peter.bokes@stuba.sk}
\affiliation{Department of Physics, Faculty of Electrical Engineering and
        Information Technology, Slovak University of Technology,
    Ilkovi\v{c}ova 3, 812 19 Bratislava, Slovak Republic}
\affiliation{ETSF, Department of Physics, University of York, Heslington, York
         YO10 5DD, United Kingdom}

\date{\today{}}

\begin{abstract}
We demonstrate that the time operator that measures the time of arrival of a quantum particle 
into chosen state can be defined as a self-adjoint quantum-mechanical operator using periodic boundary 
conditions on applied to wavefuncions in energy representation. The time becomes quantized into discreet 
eigenvalues and the eigenstates of the time operator, the stroboscopic wavepackets introduced recently 
[Phys. Rev. Lett. 101, 046402 (2008).] form orthogonal system of states. The formalism provides simple 
physical interpretation of the time-measurement process and direct construction of normalized, positive 
definite probability distribution for the quantized values of the arrival time. The average value 
of the time is equal to the phase time but in general depends on the choise of zero time eigenstate, 
whereas the uncertainity of the average is related to the traversal time and is independent of this choise.  
The general fromalism is applied to a particle tunneling through resonant tunneling barrier in 1D.
\end{abstract}


\maketitle


\section{Introduction}
\label{sec-1}

The concept of time operator in quantum mechanics is a difficult and confusing 
one\cite{Landauer94,Hauge89,Muga00}. Heisenberg formulated the time-energy uncertainity principle 
already in the early days of quantum theory, indicating through analogy with the 
position-momentum uncertainty principle, that some time operator should exist. However, 
shortly after this, Pauli argued that no such self-adjoint operator can exist~\cite{Pauli26}. 
Further development in scattering theory pursued the search for an estimates of time-scales associated 
with quantum processes, establishing the phase time, the time delay seen in the motion of 
the maximum of a wavepacket
as the relevant quantity~\cite{Wigner55,Ohmura64}. This, however, turned out to be unsatisfactory 
due to inherent ambiguity in the preparation of the wavepackets or identification of its 
maxima or other features. Several imaginative approaches, like the so called Larmor-clock 
time~\cite{Rybachenko67,Buttiker83,Sokolovski93,McKinnon95,Hauge89} or the traversal 
time~\cite{Keldysh65,Buttiker82,Landauer94} were suggested to identify the relevant time-scales.
However, no final formulation of the problem has been established nor a consensus has been reached 
if such a formalism should exist. Nonetheless, the time-scale related to tunneling is 
an extremely useful concept for relevance of many-body effects in electronic transport through 
nanostructures. This has been analyzed in the pioneering work by Jonson~\cite{Jonson80,Jonson89} where
the time scale of tunneling  was determined by the time scale of formation of the image charge, 
causing alteration of the effective tunneling barrier~\cite{Binnig84,Quek07}. More generally,
our ability to characterize the time-scales of transit or tunneling of an electron through 
a nanocontacts would be extremely helpful in understanding the importance of interactions 
in ab initio description of quantum transport~\cite{Kurth10,Mera10,Vignale09,Myohanen08,Koentopp05,Sai05,Jung07}.

Independently of these physically motivated treatments, a important step forward in understanding
the time operator, not as a self-adjoint operator but rather as a positively valued 
operator measure, has been done by Holevo~\cite{Holevo78}. Independently, Kijowski~\cite{Kijowski74} 
heuristically constructed a distribution of a time-of-arrival for a quantum particle.  
His work was later developed into the formulation of the construction of probability 
distribution of the arrival time\cite{Delgado97,Leon00,Hegerfeldt04,Hegerfeldt10}.
Most recently, even the problem of non self-adjointness of the time operator for a free quantum particle 
has been addressed by Galapon~\cite{Galapon04,Galapon05} by introducing the confined time of 
arrival operator (CTAO) within finite space using specific boundary conditions in the real space.

In the present paper, we propose an alternative formalism, the use of periodic boundary conditions
in the energy representation that leads to a family of self-adjoint time operators.
In contrast to the CTAO, arbitrary scattering potentials can be considered from the start
and the related issue of normalization of the probability distribution for times~\cite{Leon00,Hegerfeldt04}
is also resolved. The boundary conditions in the energy representation lead to formal 
quantization of the time, similarly to the situation with the CTAO.
The quantization of time is a useful mathematical tool for drawing a simple physical picture 
of the time-dynamics of the quantum particle within the orthogonal time-eigenstate basis. 
This is used for physical interpretation of the zero-time eigenstate and its relation to the conventional 
arrival-time operator~\cite{Muga00} and the 'time-of-presence' operator~\cite{Holevo78}. 
However, within the energy representation, the formalism is similar to many previous uses of the 
time-operators in the form of diferentiation 
by energy~\cite{Ohmura64,Holevo78,Pacher05,Torres-Vega07,Ordonez09,Olkhovsky09}. 
Our formalism is demonstrated on a simple example of scattering of a particle on a resonant potential in 1D.

\section{Definition and general proparties of the time operator}
\label{sec-2}

In our work we will consider a quantum particle moving along x axis, characterized 
by its Hamiltonian $\hat{H}$ assumed to have continuous spectrum, occupying a state $\ket{\phi}$. 
For this particle we introduce a family of time operators, $\hat{\tau}_\eta$, where each of them
gives the time it took for the particle to arrive into the state $\ket{\phi}$, assuming its dynamics 
has been governed by $\hat{H}$. The concept of the time operator also demands a definition of 
the zero of the time for certain state, which will be discussed after the operator is introduced. 
Different zero-time states are in one-to-one correspondence to different members of the family 
of time operators.

For the state $\ket{\phi}$, we will assume that it can be expressed as a linear 
combination of the Hamiltonian eigenstates, $\ket{\epsilon, \alpha}$, with energies from a finite interval,
the \emph{energy band} $\epsilon \in (\epsilon_0,\epsilon_0+\Delta \epsilon)$, 
\begin{equation} 
	\ket{\phi} = \sum_{\alpha} \int_{\epsilon_0}^{\epsilon_0 + \Delta \epsilon} 
	d\epsilon g_\phi(\epsilon,\alpha) \ket{\epsilon, \alpha}, 
	\label{eq-def-1}
\end{equation}
where $\alpha$ is the quantum number for degenerate states at the energy $\epsilon$. This degeneracy
arises from some other operator $\hat{A}$ (or possibly several operators) that commutes with 
the Hamiltonian. The eigenstates are chosen so as to be common eigenstates of both $\hat{A}$ 
and $\hat{H}$. The complex amplitude $g_\phi(\epsilon,\alpha)$ represents the state $\ket{\phi}$ 
in the \emph{energy-band representation}. The eigenstates are normalized to the delta function of energy, 
and the amplitude is normalized to one. This representation is not unique, another one can 
be obtained choosing a different operator $\hat{B}$ that also commutes with $\hat{H}$ 
(but not with $\hat{A}$). The representations are then related through an energy-dependent 
unitary transformation
\begin{equation} 
	\ket{\epsilon,\alpha} = \sum_\beta U_{\alpha,\beta}(\epsilon) \ket{\epsilon,\beta}. \label{eq-U}
\end{equation}

We will assume that the amplitude for the state $\ket{\phi}$ has a support within the considered energy 
band so that at the ends of the interval we have $g_\phi(\epsilon_0,\alpha) = 
g_\phi(\epsilon_0+\Delta \epsilon,\alpha)=0$. The states $g_\phi(\epsilon,\alpha)$ form a subspace 
of the Hilbert space $\mathcal{H}_{\Delta \epsilon}$ of all square-integrable states 
$g(\epsilon,\alpha)$ that are \emph{periodic} within the energy band. Extending the width of the energy band,
or considering a union of all energy bands covering the continuous spectrum of the Hamiltonian, and 
adding its possible bound states forms a complete set of states~\cite{Bokes08}. However, for the 
purpose of introducing the time operators for state $\ket{\phi}$ it is sufficient 
to consider single energy band.

We will demonstrate that in the energy-band representation, within $\mathcal{H}_{\Delta \epsilon}$, 
the self-adjoint time operator can be defined as\footnote{We use atomic units where $e=\hbar=1$.}
\begin{equation} 
	\hat{\tau}_\eta = i \delta_{\alpha,\alpha'} \Dp{~}{\epsilon} + \eta_{\alpha,\alpha'}(\epsilon). 
	\label{eq-def-2}
\end{equation}
Apart from the so-far unspecified Hermitian, energy-dependent matrix
$\eta_{\alpha,\alpha'}(\epsilon)$, and the fact that we define it only within the energy band, 
this operator has been
known for long time as the operator for time in the energy-representation. It is well known  
that it is not self-adjoint if the whole spectrum is considered~\cite{Holevo78} and 
not unique by the freedom of choice in the energy representation~\cite{Hegerfeldt10}. The former 
is removed by 
the finite energy interval and the periodic boundary conditions employed. On the other hand, the freedom
of choice of the energy representation in Eq.~\ref{eq-def-1} is related to the choice of the Hermitian matrix 
$\eta_{\alpha,\alpha'}(\epsilon)$ in the definition in Eq.~\ref{eq-def-2}. Using the unitary 
transformation, introduced in Eq.~\ref{eq-U},
$U_{\alpha,\beta}(\epsilon) = \exp\{ i \int^\epsilon d\epsilon '\nu_{\alpha,\beta}(\epsilon') \}$ 
we find a transformed time operator 
\begin{equation}
	\hat{\tau}_{\eta'} = i \delta_{\alpha,\alpha'} \Dp{~}{\epsilon} + \eta_{\alpha,\alpha'}(\epsilon) 
			 - \nu_{\alpha,\alpha'}(\epsilon).
\end{equation}
Hence, starting from a particular energy-representation and a particular choice of 
$\eta_{\alpha,\alpha'}(\epsilon)$, we can find a representation where the time operator 
is represented by the energy derivative only. This latter representation, if we had some 
rationale for choosing it independently of the time operator, could serve as the basis for definition of the 
time operator without any ambiguities. 

The first step along this line is to demand
that the time operator should commute with chosen operator(s) $\hat{A}$. 
Examples of these could be the linear or angular momentum, spin, \emph{etc.}
One of these is also the projector to the right- and left- going scattering states leading 
to $\alpha=R$ or $\alpha=L$ in 1D scattering that will be used in the next section.
This reduces the matrix $\eta_{\alpha,\alpha'}(\epsilon)$ into diagonal form 
and specifies the time operator for a processes which conserve the particular quantum number $\alpha$.
For the simplicity of notation, we will not indicate 
this fact with any additional index for the time operator $\hat{\tau}_\eta$.

Further specification of $\hat{\tau}_\eta$ is related to the choice of zero time state, as discussed 
below, but in general no unique definition of the time operator will be given.
Instead, we will accept that we deal with a family of operators of the form given 
by Eq.~\ref{eq-def-2} with $\eta_{\alpha,\alpha'}(\epsilon) = 
\delta_{\alpha,\alpha'} \eta_\alpha(\epsilon)$ and that for a specific calculations 
we need to choose one particular form.

The eigenfunctions of the time operator are
\begin{equation}
	g_{\tau_m,\alpha}(\epsilon,\alpha') = \frac{1}{\sqrt{\Delta \epsilon}}  \label{eq-tau-eig}
	e^{-i\tau_m \epsilon} e^{i \int^\epsilon d\epsilon' \eta_\alpha(\epsilon')} \delta_{\alpha,\alpha'}, 
\end{equation}
where 
\begin{equation}
	\tau_m=\frac{2\pi}{\Delta \epsilon} m, \quad m=0 \pm 1, \ldots
\end{equation}
are \emph{discreet} eigenvalues of the time operator.  Similarly to our finding, 
discreet eigenvalues were found in the construction of the confined time operator~\cite{Galapon04} 
for a free quantum particle. In both cases, the quantization is nothing fundamental and arises 
only as a result of the choice of the energy band: the size of the time-quanta could be changed 
by simply changing the width of the energy band while the final average values of the time operator 
will remain independent of this choice, which becomes obvious when using the energy representation. 
Still, choosing a wider energy band, i.e. decreasing the quantum of time, is desirable if one 
is interested in finer details of the probability distribution in time variable.

Rewriting the time eigenstates from the energy-band representation into the abstract 
form we have 
\begin{equation}
	\ket{\tau_m,\alpha} = \frac{1}{\sqrt{\Delta \epsilon}}  \label{eq-tau-state-in-energy}
	\int_{\epsilon_0}^{\epsilon_{0}+\Delta \epsilon}
        d \epsilon e^{-i\tau_m \epsilon} e^{i \int^\epsilon d\epsilon' \eta_\alpha(\epsilon')} 
	\ket{\epsilon,\alpha}. 
\end{equation}
This set of states is subspace of the stroboscopic wavepacket basis, recently introduced
for the description of open non-equilibrium electronic systems~\cite{Bokes08,Bokes09}. Here we see 
that it naturally arises as the set of eigenstates of the time operator defined on the chosen 
interval of energies.

The eigenstates with $m=0$, $\ket{\tau_0=0,\alpha}$, have eigenvalue of the 
time operator zero, i.e. this is the choice of the zero of time. 
Due to the unspecified phase $\eta_\alpha(\epsilon)$ there is a certain freedom 
in the choice of this zero time eigenstate (and hence a particular time operator).
If a particle is in one of these states, it took zero time to arrive into it.
On the other hand, for a state $\ket{\tau_m,\alpha}$ it took 
it precisely the time $\tau_m =2\pi m/(\Delta \epsilon)$ for the particle 
to arrive there from $\ket{\tau_0,\alpha}$, since from Eq.~\ref{eq-tau-state-in-energy}
we find 
\begin{equation} 
	e^{-i\hat{H} \tau_m} \ket{0,\alpha} = \ket{\tau_m,\alpha}.
\end{equation} 
Finally, for a particle in an arbitrary state within the energy band, 
$\ket{\phi}= \sum_{m,\alpha} c_{m,\alpha} 
\ket{\tau_m,\alpha}$, and $|c_{m,\alpha}|^2=|\braket{\tau_m,\alpha}{\phi}|^2$ will be the probability for the
particle to arrive into it in time $\tau_m =2\pi m/(\Delta \epsilon )$.

From the above it follows that the expectation value of the time $\tau^\phi_\eta$ in the
state $\ket{\phi}$ is given by weighting the different time eigenvalues with the probabilities 
that the relevant eigenstate is present in the state $\ket{\phi}$, 
\begin{equation} 
	\tau^\phi_\eta = \sum_{m,\alpha} |c_{m,\alpha}|^2 \tau_m
	= \bra{\phi} \hat{\tau}_\eta \ket{\phi},
\end{equation}
This is equivalent to using the form in Eq.~\ref{eq-def-2} within the energy-band representation,
\begin{eqnarray} 
	\tau_\eta^\phi = \sum_\alpha \int d \epsilon g^*_\phi(\epsilon,\alpha)  \left[
	i \Dp{~}{\epsilon} + \eta_\alpha(\epsilon) \right] g_\phi(\epsilon,\alpha). \label{eq-time-in-e}
\end{eqnarray} 
which motivates the formal definition of the time operator by the Eq.~\ref{eq-def-2}.
Clearly, the expectation value of the time operator depends on the choice of zero-time eigenstate, 
i.e. on the choice of the phases $\eta_\alpha(\epsilon)$. In contrast, for the uncertainty of this average, 
\begin{equation} 
	\Delta \tau^\phi = \sqrt{ \bra{\phi} \hat{\tau}_\eta^2 \ket{\phi} - 
				 ( \bra{\phi} \hat{\tau}_\eta \ket{\phi} )^2 }
\end{equation}
we find 
\begin{equation} 
	\Delta \tau^\phi = \sum_\alpha \int d \epsilon \Dp{~}{\epsilon} 
	\left| g_\phi(\epsilon, \alpha) \right|^2 \label{eq-tau-uncertain}
\end{equation}
which is manifestly \emph{independent} of the choice of the phases $\eta_\alpha(\epsilon)$ and hence
characteristic of the whole family of time operators.


The whole family of time operators, Eq.~\ref{eq-def-2}, fulfills the canonical commutation relation 
with the Hamiltonian $\hat{H}= \epsilon \delta_{\alpha,\alpha'}$, $[\hat{\tau},\hat{H}]=i$, if the 
latter is understood to act only on states $\ket{\phi}$ with finite support within the energy band.
(A minor technical issue that can be dealt with arises if the whole $\mathcal{H}_{\Delta \epsilon}$ 
is considered, since there the Hamiltonian is not continuous at the boundaries of the energy band.)
This commutation relation then leads automatically to the uncertainty relation for the mean square 
fluctuations in the energy and the time, $\Delta \tau \Delta H \geq 1/2$.

The argument due to Pauli~\cite{Pauli26,Muga00} regarding the non-existence of 
the self-adjoint time-operator does not apply since the boundary conditions cause 
the energy-shift operator to move the states periodically within the band.
Namely, 
using the orthogonality of the time-operators' eigenstates (Eq.~\ref{eq-tau-eig}) we 
can expand the Hamiltonian's eigenstates $\ket{\epsilon,\alpha}$, Eq.~\ref{eq-tau-state-in-energy}, 
into the former and find the identity
\begin{equation}
	\hat{H} e^{-i \epsilon' \hat{\tau}} \ket{\epsilon,\alpha} = 
	\left[ (\epsilon - \epsilon') \textrm{mod} ( \Delta \epsilon )  \right]
	e^{-i \epsilon' \hat{\tau}} \ket{\epsilon, \alpha},
\end{equation}
for $\epsilon \in (\epsilon_0, \epsilon_{0}+\Delta \epsilon)$.
On the other hand, if the periodic boundary conditions within the bands were not used, the above identity 
would not contain the modulo operation with the difference $(\epsilon-\epsilon')$ and the 
result would be that the state $e^{-i \epsilon' \hat{\tau}} \ket{\epsilon,\alpha}$ is an eigenstate
of Hamiltonian with the eigenvalue $\epsilon-\epsilon'$. Following Pauli, and in view of arbitrariness 
of $\epsilon'$,  this would be in contradiction with the existence of the lower bound on the eigenenergies. 
However, we have shown above that the use of the periodic boundary conditions removes this problem.

The here used time operator is, by its character, close to the 'time-of-presence' mentioned in the review 
by Muga and Leavens~\cite{Muga00}. However, many authors~\cite{Kijowski74,Delgado97,Hegerfeldt04,Leon00} 
prefer the concept of the arrival-time operator that gives the average value of time for a quantum 
particle to arrive at a spatial 
position $x_0$ if initially (at time $t=0$) it was in a chosen state $\ket{\psi}$. One can easily see that 
such an operator is given by $-\hat{\tau}_\eta$, with a specific choice of the phases $\eta_\alpha(\epsilon)$.
The latter is such that the zero-time eigenstates resembles the position eigenstate 
$\delta(x-x_0)$ as much as possible. For example, for a free quantum particle the energy eigenstates are
\begin{equation}
	\braket{x}{\epsilon,\alpha} = \frac{1}{\sqrt{2 \pi k}} e^{ikx}, \quad k = \sqrt{2\epsilon},
\end{equation}
and using the projection $\int d \epsilon \ket{\epsilon,\alpha}\braket{\epsilon,\alpha}{x_0}$, one finds 
$\eta_\alpha(\epsilon)=(d/d\epsilon) \sqrt{2\epsilon} x_0$. The interpretation of this time 
operator is as follows: we expand the state $\ket{\psi}$ into the arrival-time operator eigenstates,
$\ket{\psi} = \sum_m \braket{\tau_m,\alpha}{\psi} \ket{\tau_m,\alpha}$. Then 
$|\braket{\tau_m,\alpha}{\psi}|^2$ is the probability that the particle in $\ket{\psi}$ will arrive 
into the $\ket{\tau_0=0,\alpha}$ in time $-\tau_m$. Identifying the zero-time eigenstate with 
measurement device at $x=x_0$ gives the sought Kijowski probability distribution~\cite{Hegerfeldt04,Leon00}.
However, we need to stress that the arrival state, i.e. the zero-time eigenstate, can be quite different
from the position eigenstate $\delta(x-x_0)$ so that it should not be interpreted literally 
as the probability of the time of arrival into $x_0$ exactly. 

\section{Arrival time in tunneling problems}
\label{sec-2}

\begin{figure}[t]
  \includegraphics[width=8cm]{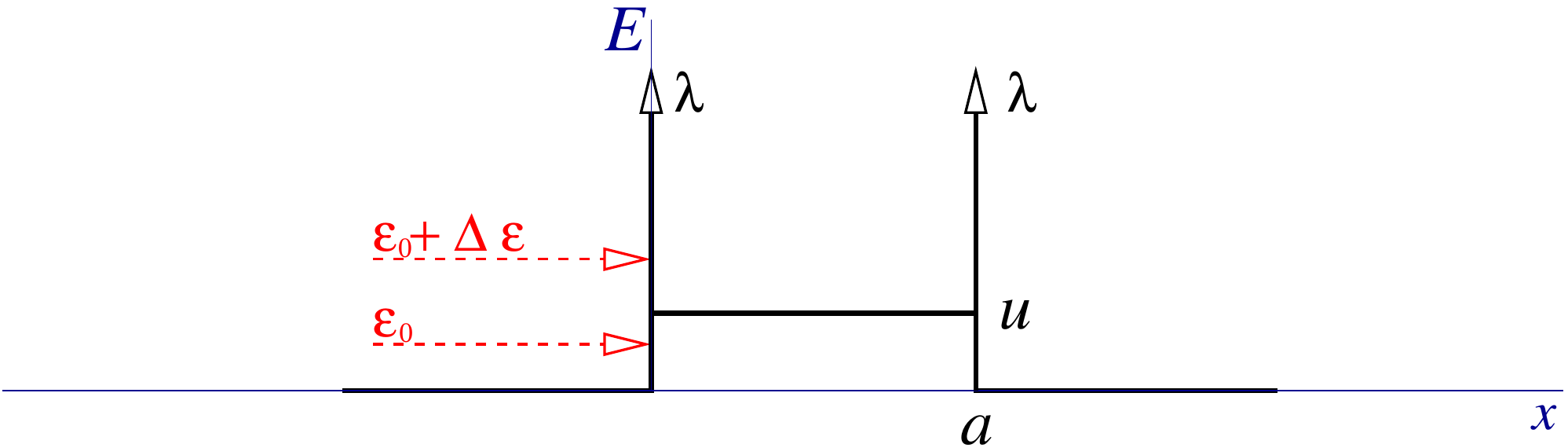}
  \caption{(color online)
	The form of the potential energy used for demonstration of the  
	tunneling time scales in 1D. The potential has two delta-shaped barriers of strength $\lambda$
	at its both ends and a constant value $u$ in between. Varying the 
	latter gives access to various transport regimes - from resonant tunneling to opaque 
	tunneling. The energy-band of the stroboscopic wavepacket representation is indicated
	with arrows. 
        } \label{fig-1}
\end{figure}

We will now demonstrate the use of the time operator for calculation 
of the tunneling time scales involved in the dynamics of quantum particle in 1D. 
For the state into which we expect the particle to arrive we initially take 
a state $\ket{\phi}$, located on the right of the tunneling barrier 
and characterized by a momentum directed away from the barrier,
\begin{equation} 
	\braket{x}{\phi} = 
	\int_{\epsilon_0}^{\epsilon_0+\Delta \epsilon} \frac{d\epsilon}{\sqrt{2\pi k}} 
	A_\phi(\epsilon) e^{ik(x-x_R)}, \quad x_R \gg 0, \label{eq-phi-x}
\end{equation}
where $x_R$ determines the average position of a particle in the state and $k=\sqrt{2\epsilon}$. 
The real amplitude $A_\phi(\epsilon)$ is a continuous, differentiable function with its support within 
the energy band. The time it takes for the particle to arrive into the state $\ket{\phi}$ from 
the left of the barrier will contain contribution of the time it took for the particle to tunnel. 

The dynamics of the particle is governed by the Hamiltonian $\hat{H}$ which asymptotically, 
for $x \rightarrow \pm \infty$, is that of a free particle. Close to the origin there is non-zero
potential energy $V(x)$. The Hamiltonian posses a continuous spectrum of doubly-degenerate 
energy-normalized right- and left- going eigenstates $\braket{x}{\psi_{\epsilon,R}}$ and 
$\braket{x}{\psi_{\epsilon,L}}$ of which we will explicitely need only the right-going ones,
\begin{equation} 
	\braket{x}{\psi_{\epsilon,R}} = \psi_{\epsilon,R}(x) = \left\{ 
		\begin{array}{ll} 
		\frac{e^{ikx}}{\sqrt{2\pi k}} + r(\epsilon) \frac{e^{-ikx}}{\sqrt{2\pi k}} & x << 0 \\
		t(\epsilon) \frac{e^{ikx}}{\sqrt{2\pi k}} & x >> 0 
		\end{array} \right. , \label{eq-psi-R}
\end{equation}
where $k=\sqrt{2\epsilon}$ and $r(\epsilon)$ and $t(\epsilon)$ are the reflexion and transmission amplitudes
respectively. 

\begin{figure}[t]
  \includegraphics[width=8cm]{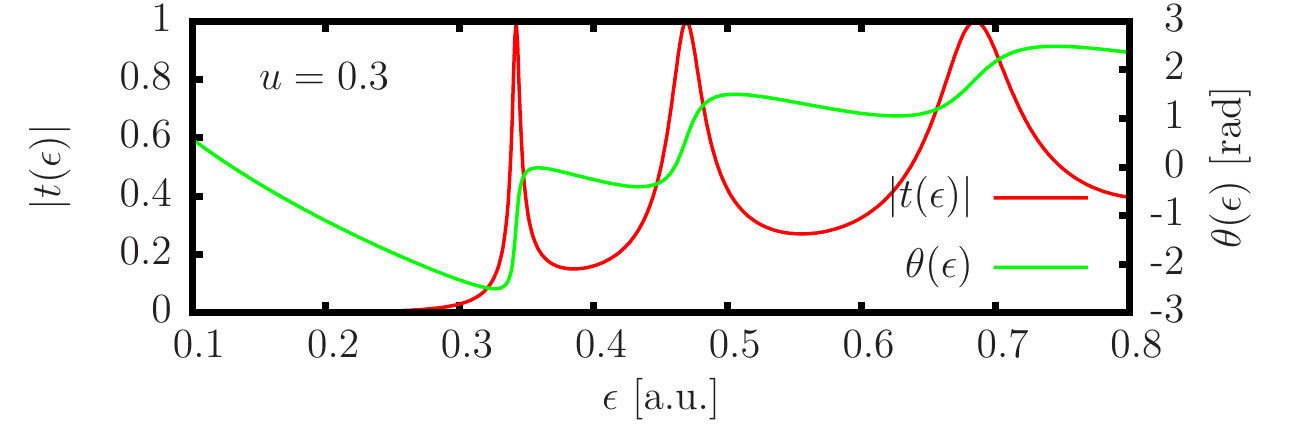}
  \includegraphics[width=8cm]{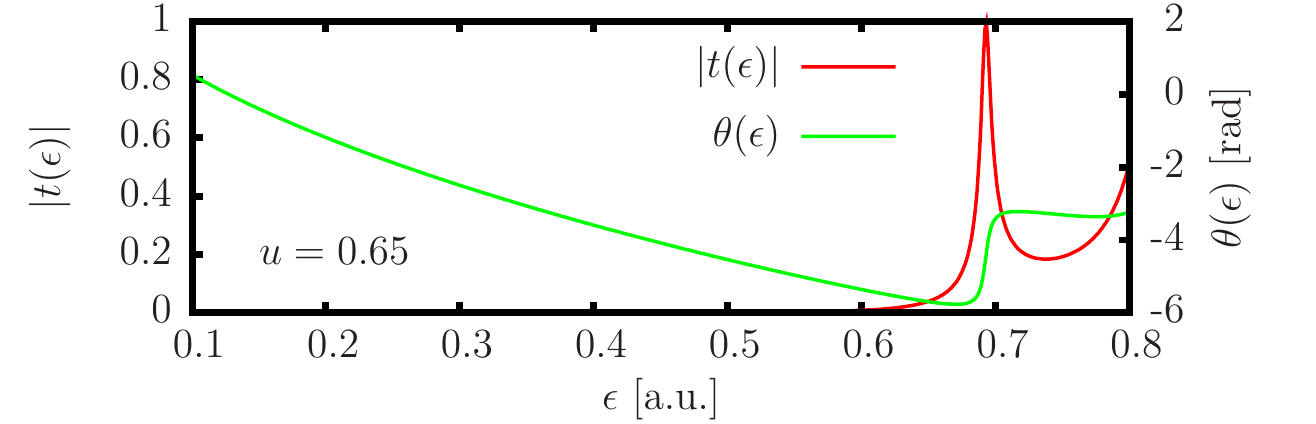}
  \caption{(color online)
	The amplitude and the phase of the transmission amplitude $t(\epsilon)$ for a scattering 
	state in the potential given in Fig.~\ref{fig-1} for $u=0.3$ (upper panel) and $u=0.65$ 
	(lower panel). Since the energy band consists of energies $\epsilon
	\in (0.2,0.6)$, the smaller $u$ corresponds to resonant transport where as the larger 
	$u$ gives dominantly opaque tunneling regime.
        } \label{fig-1b}
\end{figure}

As a illustrative example, we will consider the Hamiltonian $\hat{H} = (1/2) d^2/dx^2 + V(x)$, 
where $V(x) = \lambda ( \delta(x) + \delta(x-a) ) + (u/2)(1(x) - 1(x-a))$ where $1(x)$ is the unit-step
function, $\lambda=1$, $a=10.0$, and $v=0.1-0.65$ is a variable potential within the delta-functions
(see Fig~\ref{fig-1}). As can be inferred from the amplitude and the phase of transmission amplitude shown in 
Fig.~\ref{fig-1b}, these values offer variety of different transport regimes. It might be also 
interesting to mention that this Hamiltonian corresponds to a simple model of a perturbed monoatomic sodium 
chain~\cite{Bokes09}. For the arrival state $\phi(x)$ in Eq.~\ref{eq-phi-x} we take 
$A(\epsilon) = N_A \left[\cos^2\left((\epsilon - \epsilon_1)\pi/\epsilon_2\right) \right]$,
where $\epsilon_1 = \epsilon_0 + \Delta \epsilon/2$ and $\epsilon_2=\Delta \epsilon$, with 
$\epsilon_0=0.2$ and $\Delta \epsilon=0.4$, which produces a convenient localized state covering 
the interesting features in the transmission and its phase (Fig.~\ref{fig-1b}). The constant $N_A$ is fixed 
by the normalization of the state $\phi(x)$. The localization of the arrival state will be set 
at $x_R=100$ which for the chosen parameters will guarantee that its amplitude in the region 
of the nonzero potential is negligible.

We proceed by the selection of the time-operator. Firstly, the Hermitian matrix 
$\eta_{\alpha,\alpha'}(\epsilon)=\eta_\alpha(\epsilon)\delta_{\alpha,\alpha'}$ shall be diagonal 
in the basis of the scattering states 
so that $\alpha=R$ or $L$ for right- or left- going scattering states. Secondly, the phases 
of this matrix will form the zero-time eigenstate from the incoming scattering states 
at some initial time $T_0$, 
\begin{equation} 
	\eta_{R/L}(\epsilon) = - T_0 .
\end{equation}
If $T_0$ were $<<0$, the zero-time eigenstates would look just like wavepackets formed from incoming 
plane-waves, localized far to the left (right) of the barrier for $\alpha=R$ ($L$) respectively.
However, this only shifts its origin for the time, so that in our calculations we will simply use $T_0=0$.

To evaluate the average time of arrival into $\ket{\phi}$, we need to express this state 
in the energy representation of the above scattering states,
\begin{equation} 
	\ket{\phi} = \sum_\alpha \int d \epsilon \ket{\psi_{\epsilon,\alpha}} 
		\braket{\psi_{\epsilon,\alpha}}{\phi}, \quad \alpha = R,L
\end{equation}
For the chosen state $\ket{\phi}$, Eq.~\ref{eq-phi-x}, both $\braket{\psi_{\epsilon,R}}{\phi}$ 
and $\braket{\psi_{\epsilon,L}}{\phi}$ will be nonzero. The presence of the left-going states 
goes against our intention to characterize the tunneling time scale. The physically relevant 
state in which the time average should be calculated should consists of the right-going states only.
\begin{figure}[t]
  \includegraphics[width=8cm]{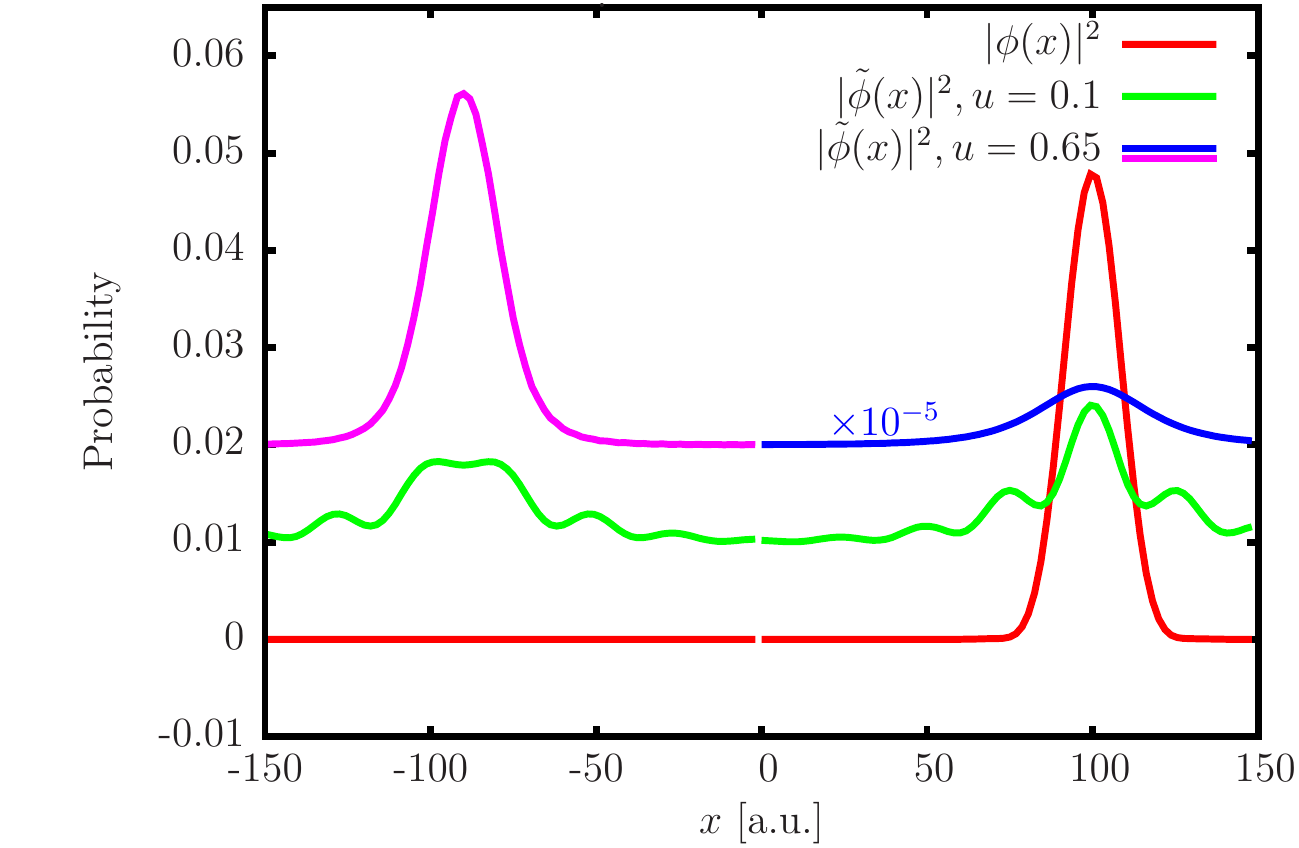}
  \caption{(color online)
	Probability densities of projected states $\tilde{\phi}(x)$ for $x_R=100$ into which the particle 
	arrives, for two different potential barriers $u$. In contrast with the unprojected state $\phi(x)$,
	the arrival states are distorted and contain significant weight on both sides off 
	the potential barrier (the amplitude for the state for $u=0.65$ on the right is $10^5$ times 
	magnified so that it is visible) localized within $x\in(0,10)$.
        } \label{fig-2}
\end{figure}
To construct the final state correctly for a particle moving from the left to the right of the barrier, 
we need to construct a state $\ket{\tilde{\phi}}$, obtained from the state $\ket{\phi}$ by a
von Neumann projection on the right-going states only, 
$\int d \epsilon \ket{\psi_{\epsilon,R}} \bra{\psi_{\epsilon,R}}$,
\begin{equation} 
	\ket{\tilde{\phi}} = \frac{1}{\sqrt{N}} \int d\epsilon  \label{eq-Neumann}
			\ket{\psi_{\epsilon,R}}  
			\braket{\psi_{\epsilon,R}}{\phi} ,
\end{equation} 
where $N$ is the normalization constant,
\begin{equation}
	N = \int d \epsilon |\braket{\psi_{\epsilon,R}}{\phi}|^2. \label{eq-norm}
\end{equation}
The state $\ket{\phi}$ for $x_R=100$ is negligibly small in the region where the states 
$\psi_R(x)$ differ 
from their asymptotic form for $x>>0$ in Eq.~\ref{eq-psi-R}, so that
\begin{equation}
	\braket{\psi_{\epsilon,R}}{\phi} = t^*(\epsilon)A_\phi(\epsilon) e^{-ik x_R}, 
		\label{eq-psiphi} 
\end{equation}
and $N=\int d \epsilon |A (\epsilon) t(\epsilon)|^2 = \langle T \rangle$
has the meaning of the average transmission probability of the particle from the left to the right. 
The unprojected $\phi(x)$ and the projected states $\tilde{\phi}(x)$
for two extremal values of the potential parameters $u$ are shown in Fig.~\ref{fig-2}. 
It should be noted, that 
unlike $\phi(x)$, the projected states are distorted and contain a non-zero amplitude of the reflected
state on the left of the barrier. This is necessary to have the total probability of arriving into 
the projected state at \emph{any} time being one.

\begin{figure}[t]
  \includegraphics[width=8cm]{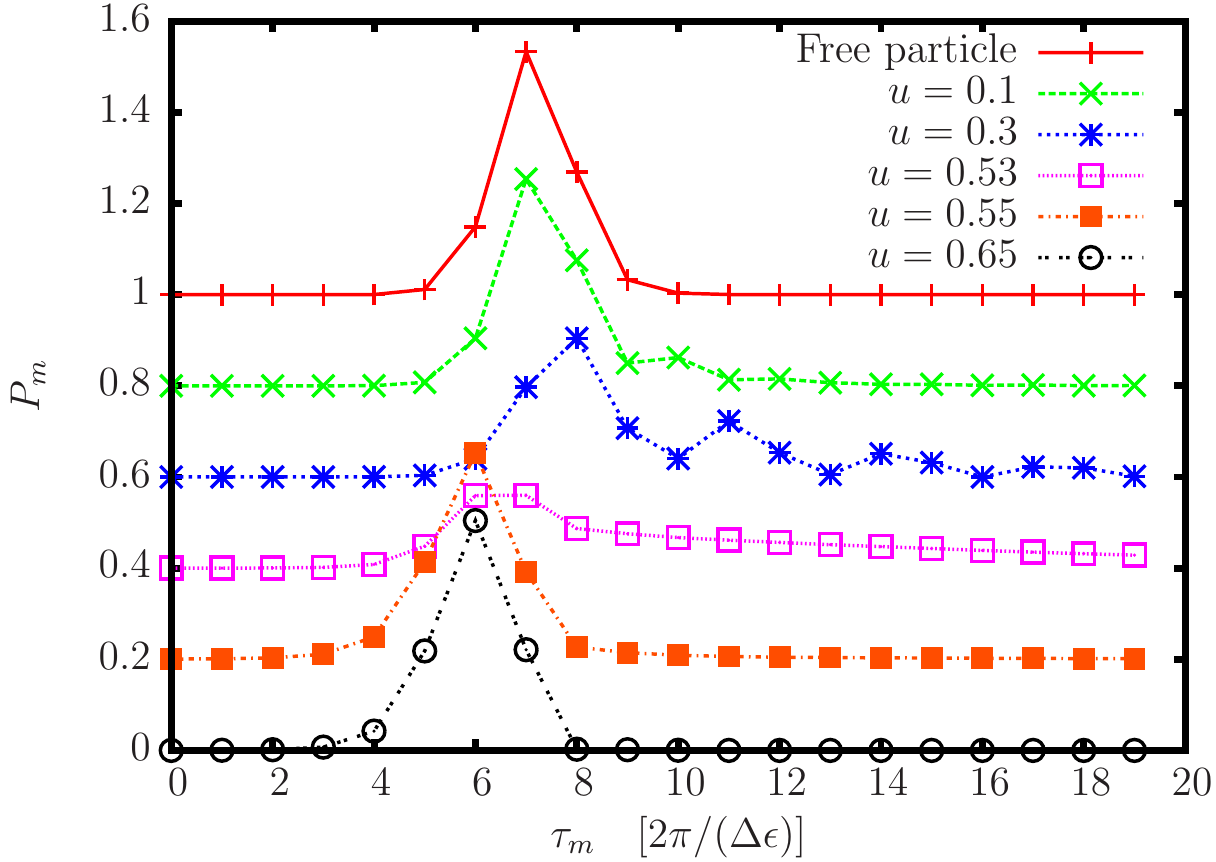}
  \caption{(color online)
	The probabilities of arrival into state $\tilde{\phi}(x)$ at time $\tau_m$ for different 
	transport regimes. In resonant transport ($u=0.3$), due to good localization of the state 
	we see several maxima giving multiple bounces between the delta-function barriers;
	increasing $u$ into tunneling regime we first see very broad distribution ($u=0.53$)
	with its width characterized by the Buttiker-Landauer time, evolving into 
	relatively narrow saturated distribution arriving earlier than the free particle (the Hartman effect).
        } \label{fig-3}
\end{figure}

According to the general treatment, the probability that the particle arrived into 
state $\ket{\tilde{\phi}}$ at time $\tau_m$ is  
\begin{eqnarray} 
	P_m &=& \left| \braket{\tau_m, R }{ \tilde{\phi}} \right|^2 \nonumber \\
	    &=& \frac{1}{\langle T \rangle} 
	    \left| \int \frac{d \epsilon'}{\sqrt{\Delta \epsilon}}  A_\phi(\epsilon) t^*(\epsilon') 
		e^{i\left[(\tau_m + T_0)\epsilon' - k x_R\right]} \right|^2 \label{eq-Pm}
\end{eqnarray}
which is positive definite and normalized to one. The example of several such probabilities 
obtained for our particular example is shown in Fig.~\ref{fig-3}. For 
the energy band above the potential between the barriers ($u=0.1, 0.3$), the particle will bounce between 
the barriers, resulting in the probability density of the arrival time into state to the right 
of the potential with several local maxima, separated by a time it takes for the particle to traverse 
the distance between the barrier twice. On the other hand, for a particle within a energy band 
below the potential barrier, no particular structure is visible. Clearly, different potentials 
lead to different distributions for arrival times and 
only the full probability distribution gives a complete picture of the time scales involved in the dynamics. 
Still for many cases two measures are most useful -- the average value and its uncertainty. 

The average time of the particle to arrive into the state $\ket{\tilde{\phi}}$ is most easily calculated 
within the energy-band representation,
\begin{eqnarray} 
	\tau_\eta^{\tilde{\phi}} 
	&=& \frac{1}{\langle T \rangle} \int d \epsilon
			|A_\phi(\epsilon) t(\epsilon)|^2 \left[ \D{\theta}{\epsilon}
				+ \frac{x_R}{v_g} - T_0 \right], \label{eq-tau-ave}
\end{eqnarray}
where $\theta(\epsilon)$ is the phase of the transmission amplitude $t(\epsilon)=|t(\epsilon)| e^{i\theta}$.
The additive term
proportional to the average position $x_R$ is the classical expression $x_R/v_g$, where $v_g=d\epsilon/dk$ 
is the group velocity of the particle outside of the barrier. The first term is the generalization
of the \emph{phase time} since for a state  within a narrow in energy band ($\Delta \epsilon \rightarrow 0$),
the average time equals the well known expression $d \theta(\epsilon)/d \epsilon$. The choice of 
the phases $\eta_{R}(\epsilon) = -T_0$ leads to a simple shift in the time, independent 
of the transmission or the Hamiltonian's potential. Keeping this form one can compare 
average times for different scattering potentials $V(x)$, localized close to the origin. In general, however, 
the energy-dependent phase leads to nontrivial change in the average time. Simple, potential-independent 
shift is found only if this energy dependence is negligible, i.e. 
$(d/d\epsilon) \eta_R(\epsilon) \Delta \epsilon \ll \eta_R(\epsilon) $.

According to Eq.~\ref{eq-tau-uncertain}, the uncertainty of the average time is given by 
\begin{equation}
	(\Delta \tau)^2
        = \int \frac{d \epsilon }{\langle T \rangle} \left( | A_\phi(\epsilon) t(\epsilon)|' \right)^2 .
\end{equation}
There are two limiting cases: (1) $(d/d\epsilon) A_\phi \gg (d/d\epsilon) |t| $ the uncertainty 
is dominated by the energy width of the state $\ket{\tilde{\phi}}$, $A_\phi(\epsilon)$, and it does 
not carry information about the time scale in the scattering, and (2) $(d/d\epsilon) A_\phi \ll 
(d/d\epsilon) |t| $ when the uncertainty is dominated by the time scale 
known previously as the traversal time~\cite{Buttiker82} and identified~\cite{Buttiker83}
as one of the Larmor clock times, $\tau^L_{z} = |t|^{-1} (d/d\epsilon) |t|$.
Most importantly, the traversal time dominates the uncertainty whenever the the energy of the state 
moves under the barrier and the transmission is principally given by 
$|t| \sim \exp\{-d \sqrt{2 (u - \epsilon)}\}$, where $d$ is the barrier width and $u$ its heigh. 
In Russian literatures it has been known as the Keldysh time\cite{Keldysh65,Yucel92} and 
is approximately given as $\Delta \tau \sim d/\kappa$ where $\kappa=\sqrt{2(u-\epsilon)}$ 
is the magnitude of the imaginary momentum under the barrier. Our result are also in agreement 
with the observations of  Yucel and Andrei that this is the time scale that should appear in 
the energy-time uncertainty principle~\cite{Yucel92}. The identification of the traversal time 
with the variance of probability density of the dwell time has been also found 
by Olkhovsky~\cite{Olkhovsky09}.

The appearance of the huge uncertainty in time, $\Delta \tau$, due to presence of the average energy 
of the state just below the barrier is demonstrated in Fig.~\ref{fig-3} for $u=0.53$, in the very 
long tail towards large times of arrival. This will have significant effect on the tunneling process
itself if the particle could interact with some additional degree of freedom within the barrier. This 
explains the appearance of this time scale in interacting tunneling 
models~\cite{Jonson80,Jonson89,Buttiker82,Yucel92}. On the other hand, increasing the potential 
barrier further, 
the uncertainty is dominated by the energy width of the state $\tilde{\phi}(x)$, and hence 
independent of $u$ (e.g.  $u=0.55$ and $u=0.65$).

Similarly to the traversal time, one can find a close correspondence between the average 
of $\hat{\tau}_\eta^2$ and the Buttiker-Landauer time~\cite{Landauer94} $\tau_T$, but this statement, 
as any expression involving the phase time, is valid only for a particular choice of the phases 
in the zero-time eigenstate. This, however, is less important in tunneling regime, where $\Delta \tau$
is the dominant contribution to the average value of $\hat{\tau}_\eta^2$.

In the conclusions, we have introduced a family of self-adjoint time operators into the framework 
of standard quantum mechanics using periodic boundary conditions used on the amplitudes within 
the energy-band representation. This representation leads to quantization of time which is a useful tool
to regularize the time-eigenstates and to draw physical interpretation of the use of this operator.
We have shown that each member of the family of time operators fulfills the canonical commutation relation 
with the particle's Hamiltonian and hence the energy-time uncertainty principle, identified its 
eigenstates are stroboscopic wavepackets, and showed how the present treatment avoids Pauli's argument. 
In the context of the tunneling of a quantum particle, 
we have shown how a positively defined distribution function of times of arrival into an arbitrary 
state is obtained. Using the latter we have obtained the average value of the time operator 
which corresponds to the phase time which, however, is specific to a particular choice of the zero-time 
eigenstate and hence the particular choice of the time operator. In contrast, its uncertainty 
is independent of this choice and in the limit of narrow energy spread of the state it is equal to
the traversal time scale of Buttiker and Landauer, or the Keldysh time. 
Our formalism confirms the role of energy derivative in energy representation as a legitimate time-operator in non-relativistic quantum mechanics and opens consistent ways for studding temporal behavior 
in many quantum-mechanical problems of interest.

The author wishes to acknowledge fruitful discussions with Martin Kon\^{o}pka.
This work was funded in part by the EU's Sixth Framework Programme through the Nanoquanta Network 
of Excellence (NMP4-CT-2004-500198) and the Slovak grant agency VEGA (project No. 1/0452/09).


\end{document}